*Original Article*

# Performance of Expansive Soil Stabilized with Bamboo Charcoal, Quarry Dust, and Lime for Use as Road Subgrade Material

Essizewa Essowedeou Agate[1], Nyomboi Timothy[2], Ambassah O. Nathaniel[3], Ines Ngassam[4]

[1]*Department of Civil Engineering, Pan African University Institute for Basic Sciences, Technology and Innovation Hosted at Jomo Kenyatta University of Agriculture and Technology, Kenya.*
[2]*Kenya Urban Roads Authority, Kenya.*
[3]*The University of Buea, Cameroun.*
[4]*The University of Cape Town, South Africa.*

[1]*Corresponding Author : agateaugustin@gmail.com*

***Abstract*** *- Expansive soils such as Black Cotton Soils (BCS) present significant challenges for road subgrade construction due to their high plasticity, swelling potential, and low strength. This study explores a triphasic stabilization method using Bamboo Charcoal (BC), Quarry Dust (QD), and Lime (L) to enhance the engineering properties of BCS for rural road applications. Initial soil characterization involved standard tests, including Atterberg limits, compaction, and Californian Bearing Ratio (CBR) assessments. The soil was treated with varying BC proportions (5% to 35% at 5% intervals) in the initial phase, leading to a progressive reduction in the Plasticity Index (PI) and swell index and an enhancement in the CBR up to 20% BC content. This further resulted in a soaked CBR value of 2.7%. In the second phase, additional treatment combined with BC and QD, incorporating diverse QD proportions (4% to 24%) relative to the optimal BC content. This further improved the CBR to 7.7% at 12% QD, but the PI exhibited a non-linear trend. Finally, 5% lime was introduced in the final phase. This minimized the PI to 11.2% and significantly increased the CBR to 19%. The optimal combination of 20% BC, 12% QD, and 5% Lime achieved optimal plasticity, compaction, and strength characteristics, demonstrating the viability of this approach for transforming BCS into a sustainable and cost-effective alternative for rural road subgrade construction.*

***Keywords*** *- Bamboo charcoal, Black Cotton Soil, Lime, Quarry Dust, Road subgrade.*

## 1. Introduction

Civil engineering, positioned at the forefront of innovation and sustainability, confronts persistent challenges posed by expansive soils, commonly known as “black cotton soils,” characterized by high plasticity and clay content. These soils, prevalent globally, pose serious threats to infrastructure, particularly in road and building construction, due to their tendency to undergo deformations, leading to differential settlement, cracking, and even collapse [1].

In many developing countries, notably in Africa, issues such as inadequate road networks and persistent traffic congestion persist, surpassing the inherent capabilities of natural soils. Traditional soil stabilization methods often involve costly materials like cement, contributing to financial and environmental concerns due to cement production’s energy-intensive and greenhouse gas-emitting nature. A ton of cement produced equals 0.8 tons of carbon dioxide released into the atmosphere, according to [2]. This necessitates exploration into alternative, cost-effective, and eco-friendly stabilizing materials.

In civil engineering, soil stabilization is critical to enhance soil strength and resistance to water softening. Soil stabilization, which involves adding materials to improve soil properties, offers mechanical and chemical approaches [3-5]. Traditional stabilizers like cement, lime, and bitumen have been extensively studied and well-established in the field [6, 7].

Specifically, lime has been utilized extensively to enhance the engineering properties of expansive soils such as Black Cotton Soil (BCS) by reducing their plasticity through cation exchange, pozzolanic reactions, and carbonation processes, collectively reducing plasticity and increasing strength by lowering the Liquid Limit (LL), raising the Plastic Limit (PL), and consequently decreasing the Plasticity Index (PI) as reported by many researchers [8, 9].

While existing literature provides valuable insights into using conventional stabilizers, a more thorough review reveals a persistent need to address the specific challenges of rural road subgrade construction using alternative, sustainable materials [10]. Recent research explores the potential of industrial and waste materials as soil stabilizers, driven by the need to reduce project costs and environmental impact. Prominent examples, including rice husk ash, corn cob ash, sugarcane bagasse ash, quarry dust, fly ash, and blast furnace slag [11-16], have demonstrated their effectiveness in enhancing construction practices.

Natural elements, mainly bamboo, have gained attention due to their sustainability potential, with bamboo charcoal, a by-product of heating bamboo in the absence of oxygen, standing out for its unique adsorption capabilities due to its micropore structure [17, 18]. Quarry dust, a residue from rock processing, has proven readily available and cost-effective, emerging as an eco-friendly alternative [19].

Studies, such as the one by [14], have investigated the effectiveness of quarry dust in enhancing the engineering properties of Black Cotton Soil (BCS), demonstrating significant improvements in California Bearing Ratio (CBR) from 1.75% to 7.05% with the addition of 40% quarry dust, and compaction parameters. The study highlighted the addition of quarry dust to black cotton soil, which reduces its expanding nature. Similarly, [20] explored the impact of quarry dust as a stabilizer for expansive soils, revealing improvements in specific gravity, liquid limit, plasticity index, compaction properties, and a decrease in potential swelling.

In the realm of natural elements, [21] aimed to improve the bearing capacity of clay soil by adding bamboo charcoal powder, demonstrating increased CBR values with higher levels of bamboo charcoal powder. The highest increase in the CBR value of the original clay is the addition of 15% bamboo charcoal powder in 60 blows, with a percentage increase of 82.87%. Another study by [5] investigated the use of activated carbon to enhance the geotechnical properties of soft clay soil, showing improvements in specific gravity, plasticity index, and CBR values with increased activated carbon content.

According to [5] and [21], significant carbon content in bamboo charcoal powder can increase the CBR value. Additionally, according to [29], the ideal lime content in clay soil raised the Optimum Moisture Content (OMC) while lowering the Maximum Dry Density (MDD) and Plasticity Index (PI).

Collectively, these studies showcase the significant improvement in the physical and mechanical properties of expansive soils achieved by incorporating materials such as bamboo charcoal, quarry dust, and lime. This provides valuable insights for soil stabilization in civil engineering applications and underscores the importance of seeking alternative, sustainable methods for enhancing soil strength and counteracting water softening. The research aligns with recent studies investigating the effectiveness of alternative materials, highlighting their potential to improve the engineering properties of expansive soils.

Within this context, our analysis focuses on a novel triphasic stabilization approach, utilizing Bamboo Charcoal (BC), Quarry Dust (QD), and Lime (L). This approach transforms Black Cotton Soils (BCS) into a sustainable and cost-effective alternative for rural road subgrade construction.

By delving into the engineering properties of BCS, the study explicitly addresses challenges related to high plasticity, swelling potential, and low strength, positioning the triphasic approach as a promising solution that contributes to the advancement of effective, cost-efficient, and environmentally friendly soil stabilization practices for expansive soils in subgrade construction.

## 2. Materials and Methods

### *2.1. Materials*

The materials utilized in this study comprise Black Cotton Soil (BCS), Bamboo Charcoal (BC), Quarry Dust (QD), hydrated Lime (L), and water.

Black cotton soil samples were systematically collected from various depths and locations near JKUAT, Juja, ensuring the acquisition of a representative sample reflecting local soil conditions. Subsequently, the accumulated soil underwent air-drying in the civil engineering laboratory at JKUAT before being further processed. Quarry dust, obtained as a by-product of stone crushing from Ndarugu quarries in Kenya, was introduced as an additional material. The collected sample was dried and sieved using the BS 425 μm sieve before use according to [26].

Bamboo waste, procured from Nandi County in Kenya, underwent air-drying followed by carbonization with minimized oxygen exposure. The resultant charred bamboo products were cooled to room temperature, crushed, and sieved using the BS 425 μm sieve before their incorporation into the study (Figure 1), as per guidelines [26]. The lime used in this study is hydrated lime sourced from suppliers in Kenya.

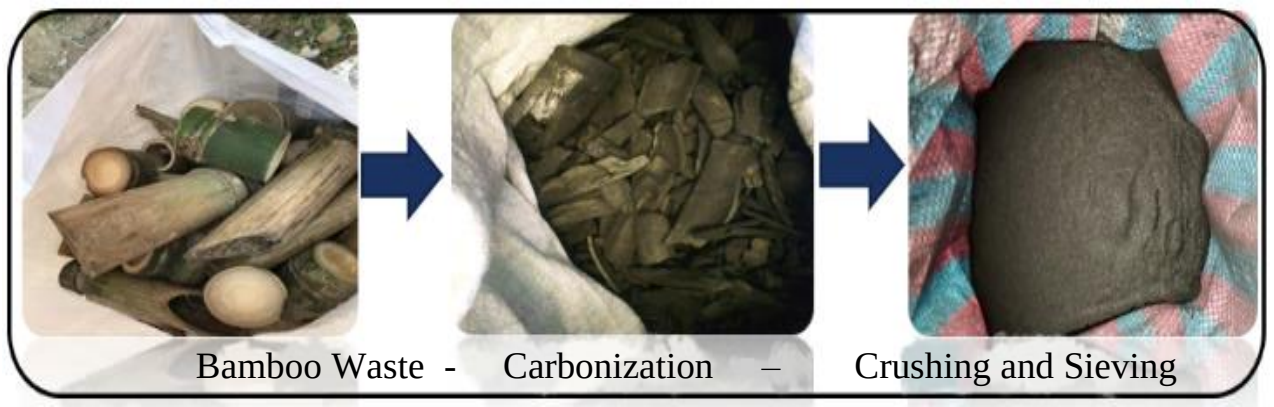

**Fig. 1 Bamboo Charcoal production process**

Potable water, sourced from Jomo Kenyatta University of Agriculture and Technology (JKUAT), served as the medium for mixing and curing the samples. This comprehensive selection of materials was deliberately chosen to faithfully replicate local soil conditions while incorporating sustainable elements into the research framework. Using indigenous materials and environmentally conscious practices, such as bamboo waste repurposing, contributes to the study's ecological considerations.

### *2.2. Methods*

Chemical composition analysis was performed on the Black Cotton Soil sample, Bamboo Charcoal (BC), Quarry Dust (QD), and Lime (L) using X-Ray Fluorescence (XRF) tests at the Ministry of Mining and Petroleum in Kenya (Table 2), and a Scanning Electron Microscopy (SEM) test was performed at the University of Cape Town, in South Africa, to understand the microstructure of untreated and treated soil (Figure 18).

Following British Standard 1377 [26], several characterization tests, such as those measuring particle size distribution, specific gravity, Atterberg limits, free-swell, compaction, soaked California Bearing Ratio (CBR), and Scanning Electron Microscopy (SEM) tests, were carried out to assess the mechanical and physical characteristics of the natural soil sample.

For the Atterberg Limits Test, a cone penetrometer, spatula, balance, and oven were utilized. The Compaction Test involved using a Standard Proctor compaction mould and hammer, balance, mixing tools (spoon, bowl), moisture cans, and graduated cylinders. The CBR Test employed a California Bearing Ratio (CBR) testing machine, mould assembly for the CBR test, swell plate, surcharge weight, dial gauges for deformation measurements, load ring, penetration piston, balance, graduated cylinders, and compaction tools.

The soil underwent treatment in three phases (Table 3): Bamboo Charcoal (BC) Treatment: The natural soil sample received treatment with varying amounts of Bamboo Charcoal (BC) at 5%, 10%, 15%, 20%, 25%, 30%, and 35% of the dry weight of the soil sample. The optimal BC content required to meet targeted specifications was then determined (Figures 3 - 7).

Optimal BC and Quarry Dust (QD) Treatment: Quarry Dust (QD) was gradually added at 4%, 8%, 12%, 16%, 20%, and 24%, starting from the optimum Bamboo Charcoal content to maximize the performance of their combination (Figures 8 - 12).

Optimal BC+QD and Lime (L) Treatment: Lime (L) was introduced at varying concentrations of 2%, 4%, and 5%. The treated soil samples underwent subsequent testing, including Atterberg limits, compaction, and CBR assessments, following British Standard 1924 for stabilized samples [27] (Figures 13 - 17).

## 3. Results and Discussion

### *3.1. Characterization of Black Cotton Soil (BCS)*

The natural soil sample's particle size distribution is shown in Figure 2, and Table 1 lists its additional engineering characteristics. The Black Cotton Soil (BCS) used in this study was classified as A-7-6 using the AASHTO system, indicating predominantly fine-grained soil with high clay content [28].

Its high Plasticity Index (PI) of 23.14 (LL = 52.56%, PL = 29.42%), its significant swell index (105%), indicating its susceptibility to substantial expansion when exposed to moisture, and significant linear shrinkage of 14.86% confirmed its expansive nature. It was classified as unsuitable for direct use as subgrade material without proper stabilization [27, 28, 22]. According to [23], a free swell value exceeding 50% is categorized as severe swelling, and a value surpassing 100% is labelled as very severe swelling.

The dominant presence of silica (76.136%) aligns with the clay mineralogy. In comparison, the relatively high iron oxide content (8.887%) suggests potential for shrink-swell behaviour due to swelling clay minerals like smectite and illite [1, 24] (Table 1). Additionally, the low packing efficiency indicated by a specific gravity of 2.23 further emphasized the need for stabilization to improve the soil's suitability for rapid subgrade applications.

**Table 1. Engineering properties of the natural Black Cotton Soil**

| Properties | Proportion/Value |
|---|---|
| Clay | 62% |
| Silt | 28% |
| Sand | 8% |
| Gravel | 2% |
| % Passing through BS Sieve 75μ | 90% |
| Colour Observation | Dark grey |
| Natural Moisture Content | 8.28% |
| Classification (AASHTO) | A-7-6 |
| Specific Gravity (Gs) | 2.23 |
| Free Swelling Index (FSI) | 105% |
| Liquid Limit (LL) | 52.26% |
| Plastic Limit (PL) | 29.42% |
| Plasticity Index (PI) | 23.14% |
| Linear Shrinkage (Ls) | 14.86% |
| Maximum Dry Density (MDD) | 1.388g/cm³ |
| Optimum Moisture Content (OMC) | 25.5% |
| CBR (4 Days Soaked) | 1.42% |

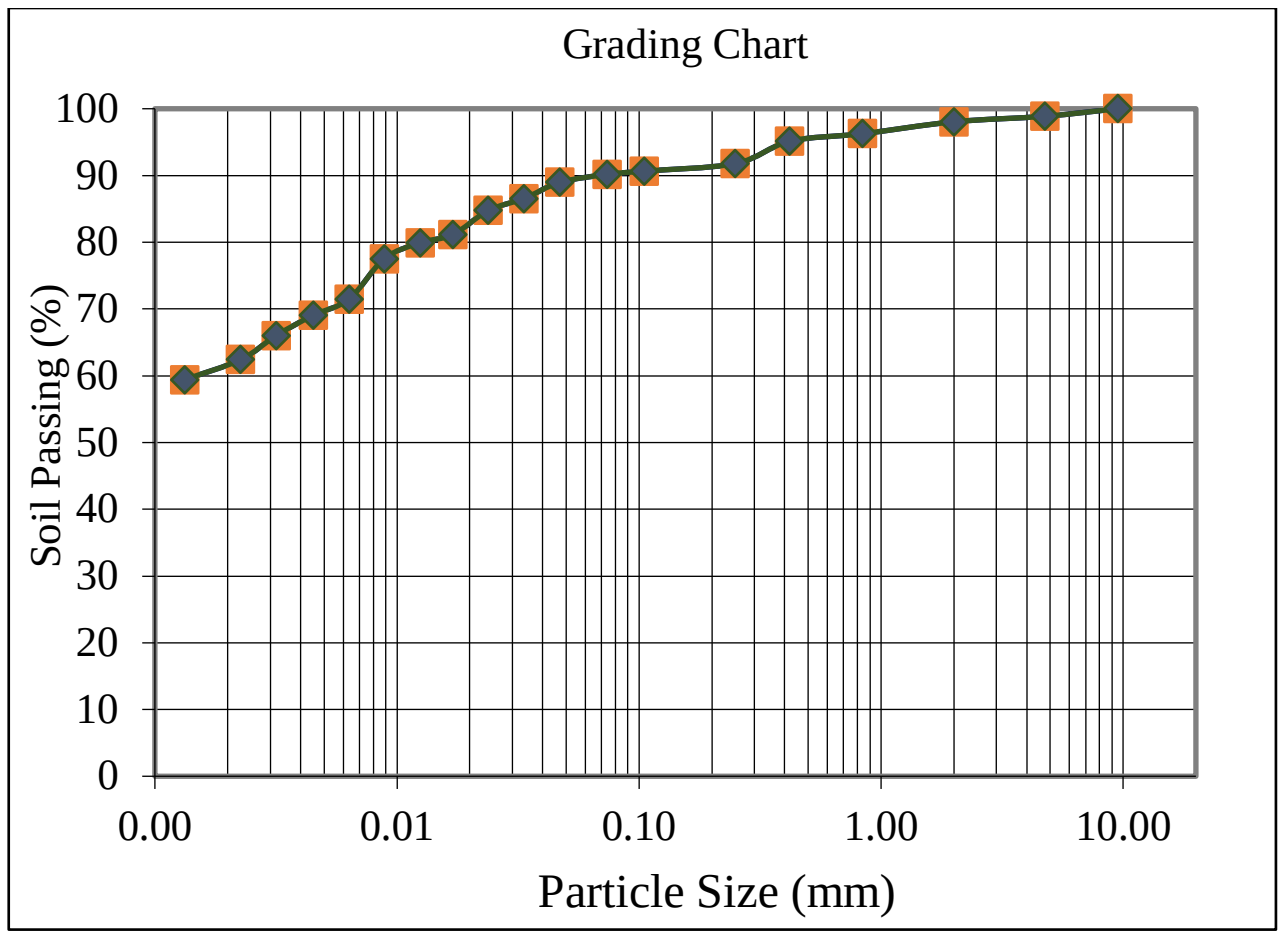

**Fig. 2 Grain size distribution curve of the soil**

**Table 2. Chemical composition of the BCS, BC, QD, and lime**

| Element | Percentage (%) | | | |
|---|---|---|---|---|
| | BCS | BC | QD | Lime |
| MgO | 0.00 | 10.184 | 0.00 | 3.393 |
| $Al_2O_3$ | 11.183 | 3.836 | 12.086 | 2.550 |
| $SiO_2$ | 76.136 | 32.502 | 72.896 | -- |
| Fe | 8.887 | 6.019 | 5.945 | 0.300 |
| CaO | 1.479 | 7.118 | 1.603 | 92.262 |
| $K_2O$ | 0.511 | 26.543 | -- | -- |
| Ti | 0.973 | 1.017 | -- | 0.045 |
| Mn | 0.540 | 0.268 | -- | 0.027 |
| Cr | -- | 0.991 | -- | -- |
| $P_2O_5$ | -- | 2.453 | 6.456 | 0.696 |
| S | -- | 1.853 | -- | 0.647 |
| Cl | -- | 6.239 | -- | -- |
| Cu | -- | 0.242 | -- | -- |

**Table 3. Black Cotton Soil treatment mix design**

| Proportions of Mix Design (%) | | |
|---|---|---|
| 100%BCS + Varying BC | 100%BCS + 20%BC + Varying QD | 100%BCS + 20%BC + Varying Lime |
| 0 | 0 | 0 |
| 5 | 4 | 2 |
| 10 | 8 | 4 |
| 15 | 12 | 5 |
| 20 | 16 | - |
| 25 | 20 | - |
| 30 | 24 | - |
| 35 | - | - |

### *3.2. Stabilization of the Black Cotton Soil with Bamboo Charcoal*

*3.2.1. Consistency Limits and Workability*

Adding Bamboo Charcoal (BC) from 0% to 35% significantly improved the workability and reduced the plasticity of the Black Cotton Soil. The initial high Plasticity Index (PI) of 23.14% steadily decreased with increasing BC content, reaching a minimum of 20.77% at 35% BC (Figure 3). Similarly, both Liquid Limit (LL) and Plastic Limit (PL) exhibited downward trends, further highlighting the improved workability and reduced susceptibility to moisture-induced deformations (Figure 3). This emphasizes the importance of selecting an optimal BC content to balance minimizing plasticity and maintaining favourable LL and PL.

*3.2.2. Shrinkage Limits and Volume Stability*

Analysis of the Black Cotton Soil treated with BC revealed a consistent reduction in shrinkage limit as BC content increased from 0% to 35%. The initial shrinkage limit of 14.89% dropped progressively to a minimum of 10.57% at 35% BC (Figure 4). This trend can be attributed to the porous structure of BC, which enhances overall soil porosity and facilitates better water penetration and drainage [7, 9]. Consequently, the soil experiences less shrinkage during drying, demonstrating the positive impact of BC's porosity on its stability and moisture retention properties [10].

*3.2.3. Free-Swell Index (FSI)*

The untreated Black Cotton Soil (BCS) sample displayed a high % free swell of 105% (Table 1), indicating its expansive nature. The free swell gradually decreased as the Bamboo Charcoal (BC) content increased from 0% to 35% (Figure 5). Specifically, at 35% BC content, the free swell reduced to 66%.

This trend suggests that BC addition led to a reduction in expansiveness. As stated by [17, 18], the porous structure of BC likely contributed to improved soil porosity and moisture retention properties, reducing susceptibility to moisture-induced swelling.

*3.2.4. Compaction Characteristics*

Compaction tests provided insights into the soil's packing efficiency and behaviour changes with BC addition. The decrease in Maximum Dry Density (MDD) from 1.388 g/cm³ to 1.200 g/cm³ and the simultaneous increase in Optimum Moisture Content (OMC) from 25.5% to 30.0% suggest alterations in the compacted state as reported by [21] (Figure 6). The lighter, less compacted nature of the soil with higher BC content implies a need for adjusting construction practices and material usage for subgrade applications.

*3.2.5. Strength Improvement (CBR for Four Days Soaked)*

California Bearing Ratio (CBR) results (Figure 7) showcase the significant enhancement in the Black Cotton Soil's strength characteristics upon stabilization with BC. The initial CBR value of 1.42% increased substantially to a peak of 2.7% at the optimal BC content of 20%. This suggests that the stabilized soil possesses improved load-bearing capacity and resistance to deformation under applied loads.

According to [3] and [21], significant carbon content in bamboo charcoal powder can increase the CBR value. Beyond 20% BC, a decline in CBR values indicates diminishing returns and emphasizes the criticality of selecting the appropriate BC content for optimal soil stabilization.

### *3.3. Stabilization of the Black Cotton Soil, Bamboo Charcoal, and Quarry Dust*

In this phase of the study, Quarry Dust (QD) was incorporated to enhance the mechanical properties of the soil and attain specified strength parameters through incremental adjustments of 4%.

#### *3.3.1. Consistency Limits and Workability*

Building upon the improved workability achieved with Bamboo Charcoal (BC) in Phase 1, this phase incorporated Quarry Dust (QD) to fine-tune the plasticity characteristics further. The results (Figure 8) demonstrated a nuanced response to the QD introduction. Notably, the Liquid Limit (LL) exhibited a progressive decrease from 66% to 62% as QD content increased, reaching its lowest at 12% QD. The Plastic Limit (PL) initially decreased to 8% QD but rose slightly to 12% QD. Consequently, the Plasticity Index (PI) displayed a non-linear trend, showcasing the interplay between BC and QD. Interestingly, the optimal combination for workability was observed at 24% QD, resulting in a significant reduction in PI from 26.1% (Soil + BC) to 20.86% (BCS + BC + 24% QD). This improvement underscores the synergistic effect of BC and QD in enhancing workability and reducing susceptibility to moisture-induced deformations (Figure 8).

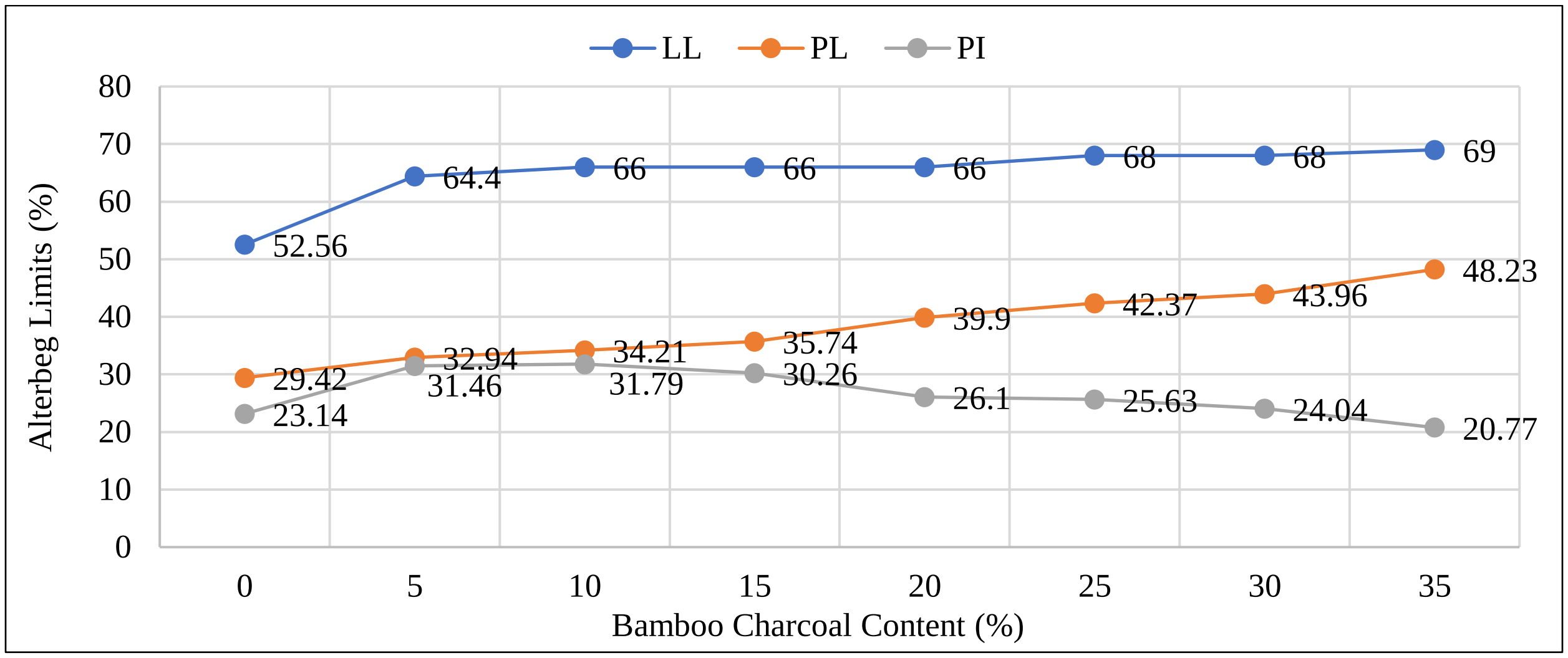

**Fig. 3 Effect of Bamboo Charcoal addition on Atterberg limits of the Black Cotton Soil**

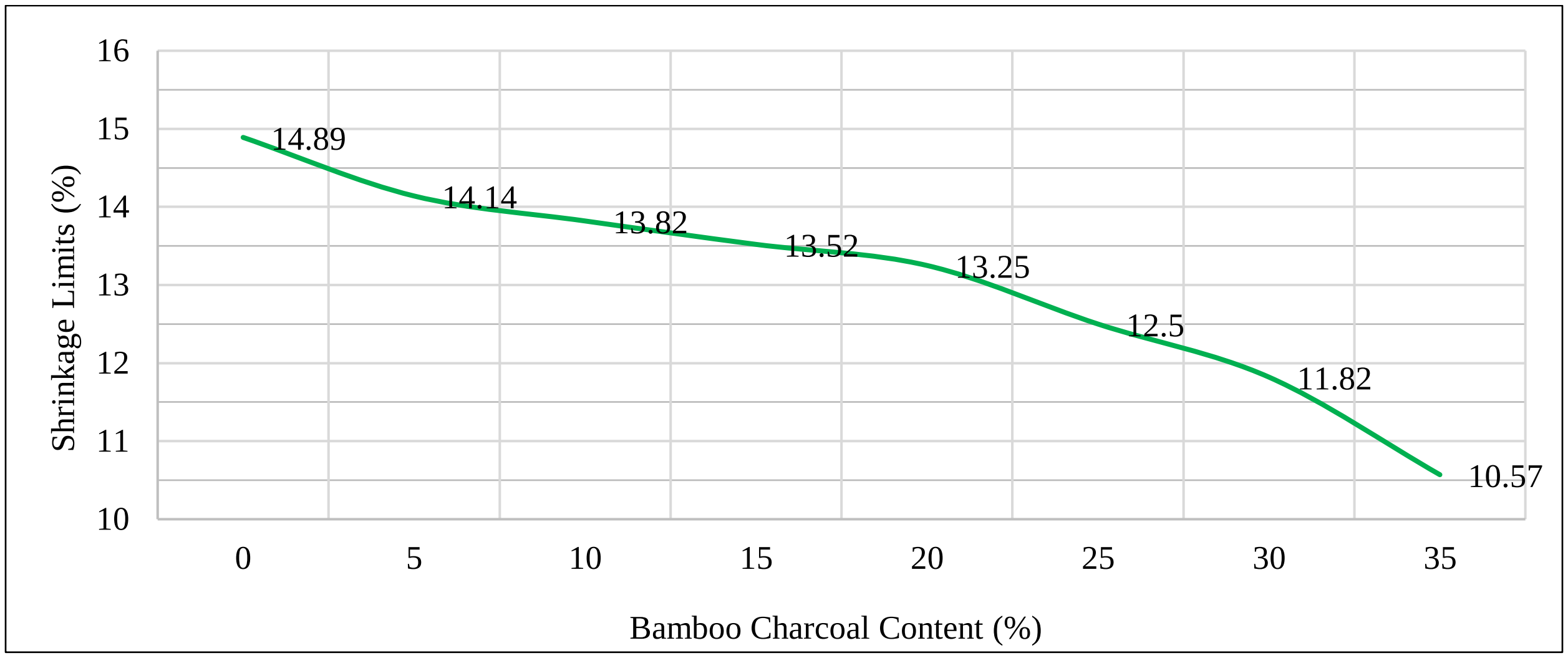

**Fig. 4 Effect of Bamboo Charcoal addition on shrinkage of the Black Cotton Soil**

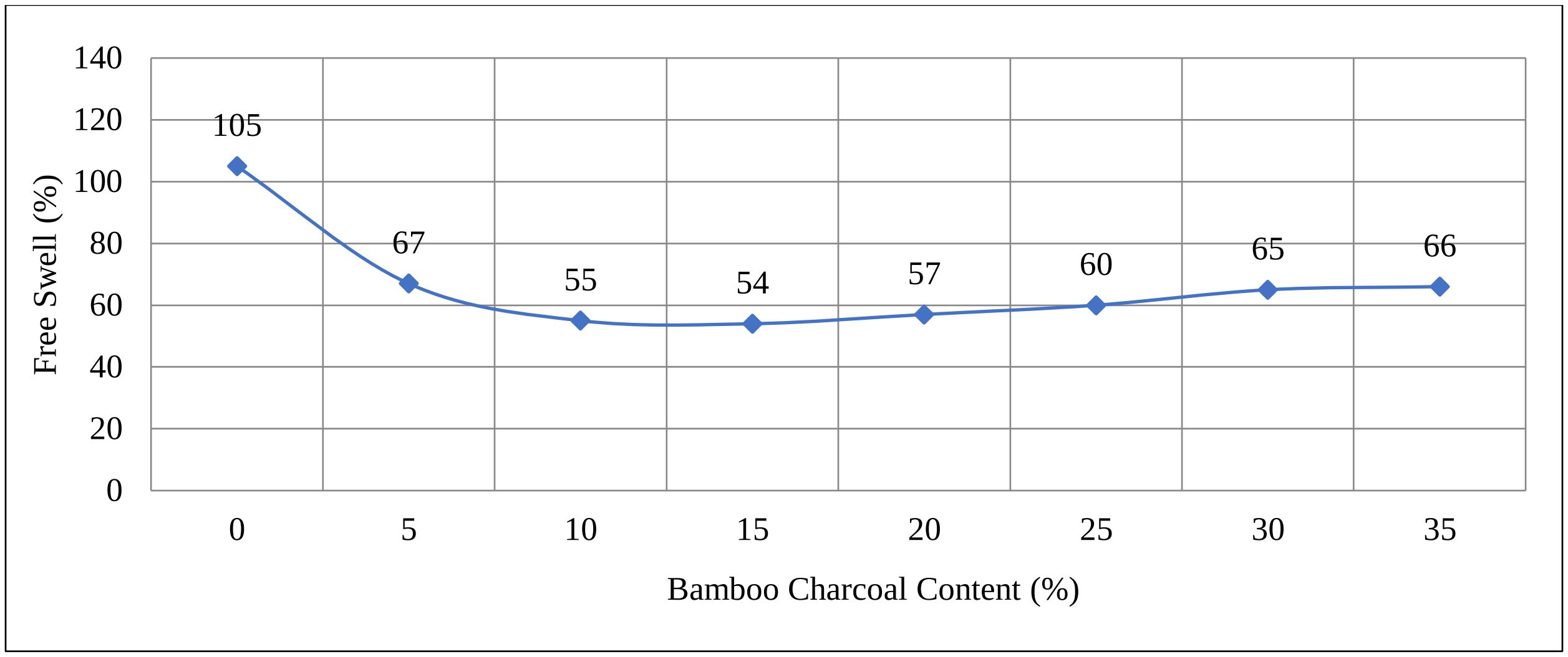

**Fig. 5 Effect of Bamboo Charcoal addition on the Free Swell of the Black Cotton Soil**

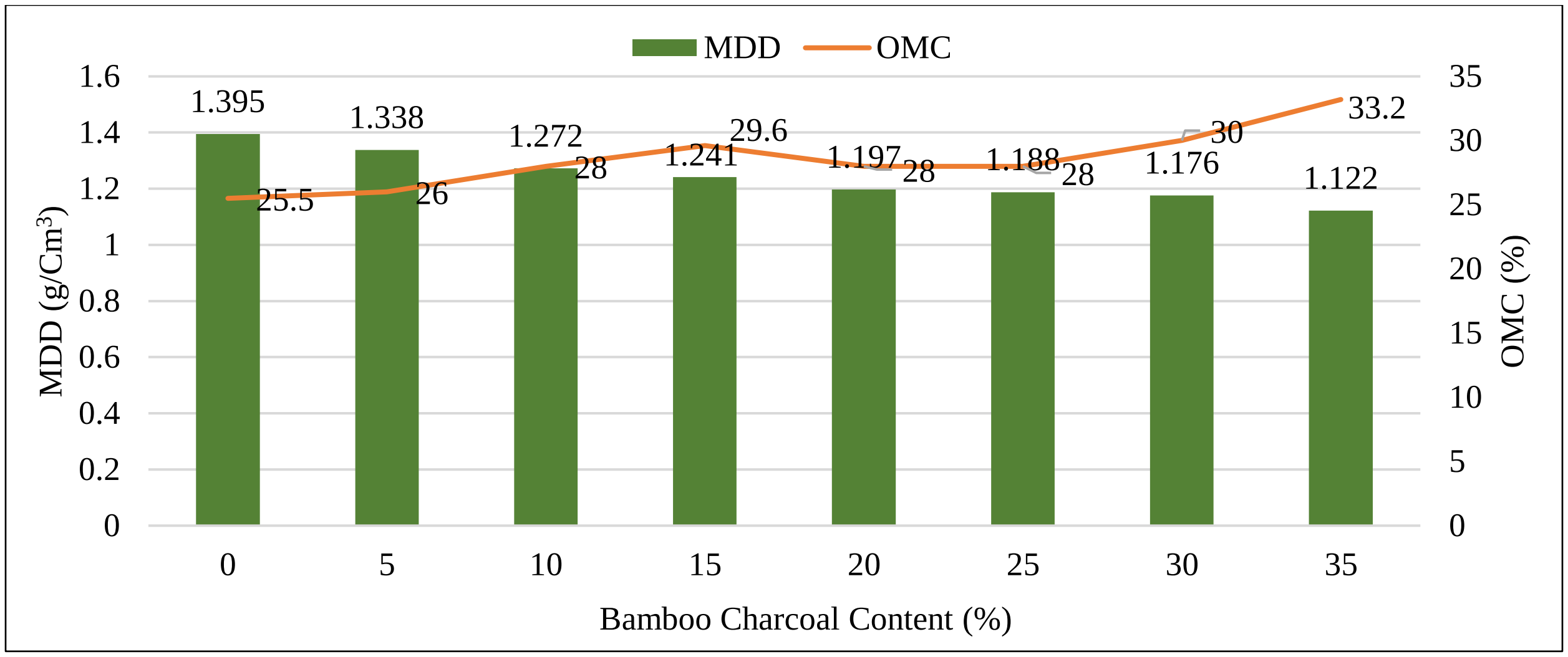

**Fig. 6 Effect of Bamboo Charcoal addition on MDD and OMC of the Black Cotton Soil**

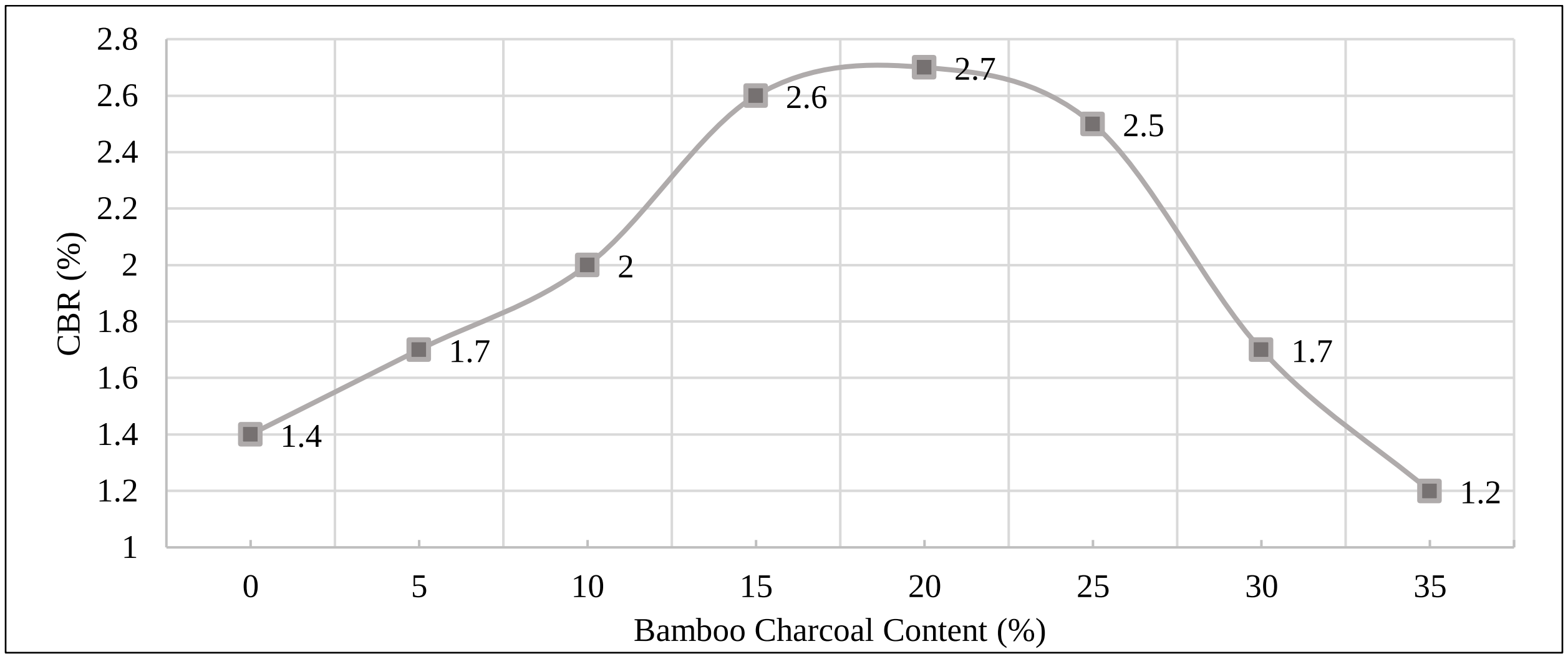

**Fig. 7 Effect of Bamboo Charcoal addition on CBR value of the Black Cotton Soil**

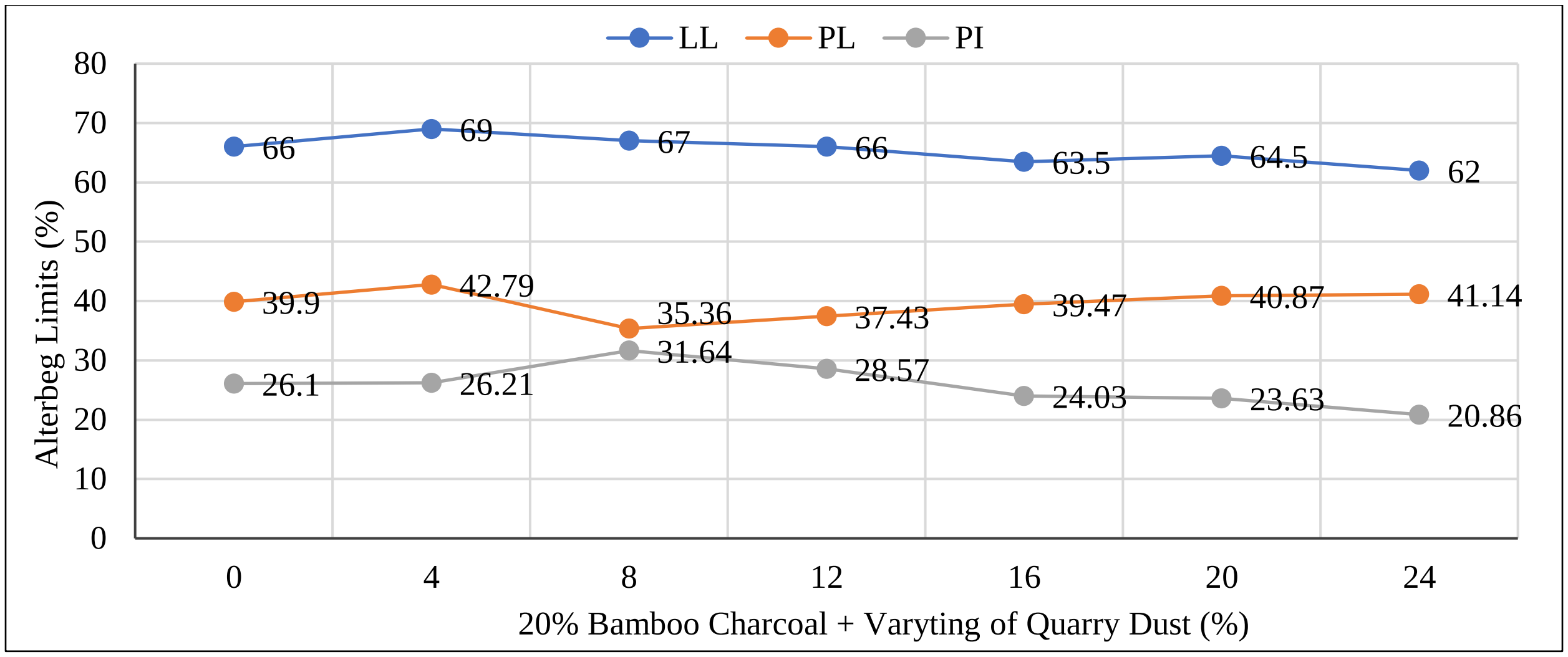

**Fig. 8 Effect of BC-QD addition on Atterberg limits of the Black Cotton Soil**

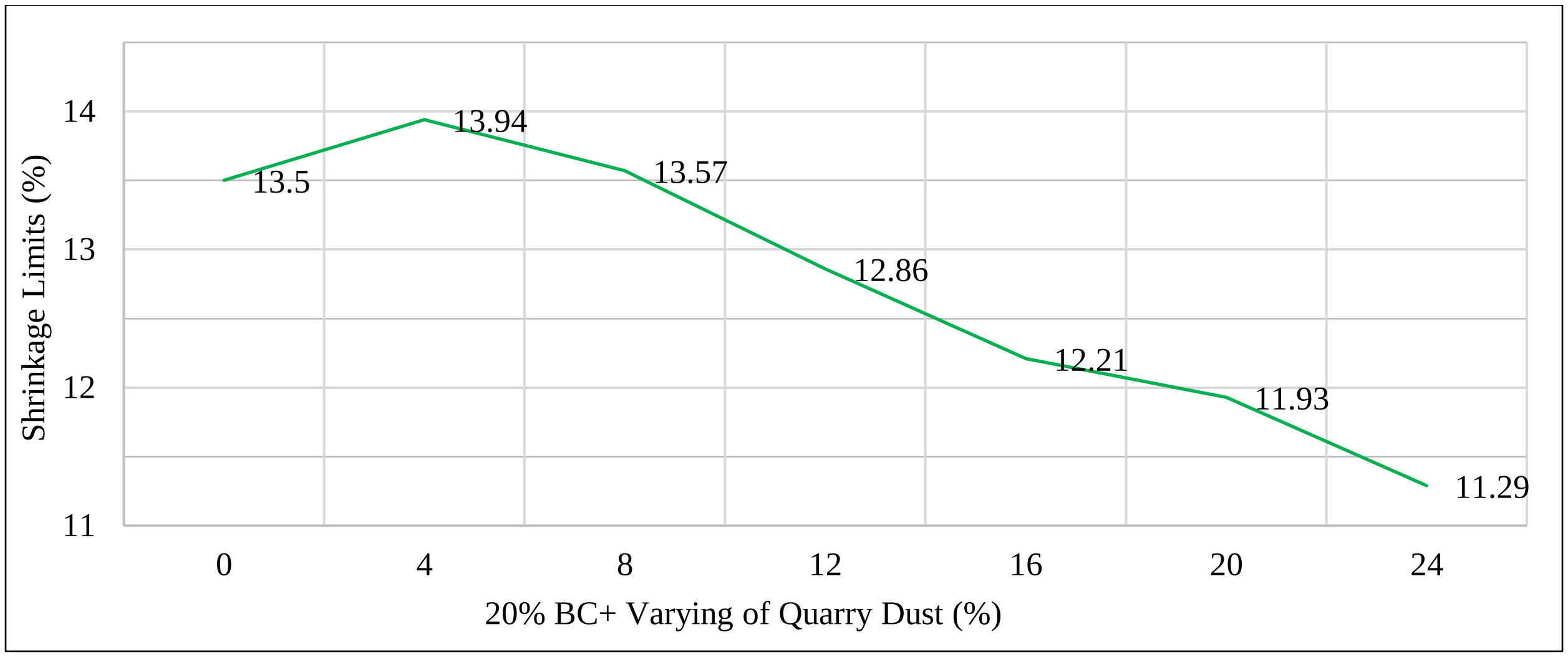

**Fig. 9 Effect of BC-QD addition on shrinkage limit of the Black Cotton Soil**

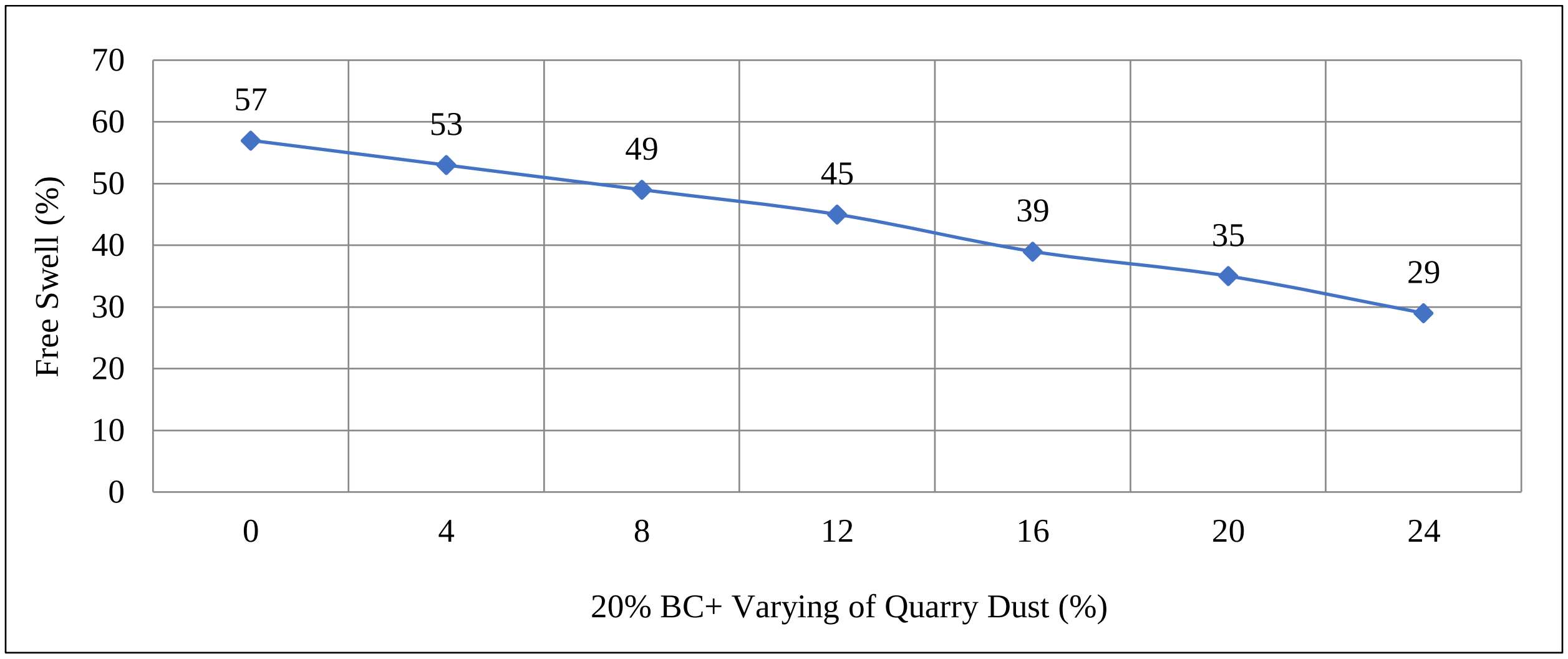

**Fig. 10 Effect of BC-QD addition on the Free Swell of the Black Cotton Soil**

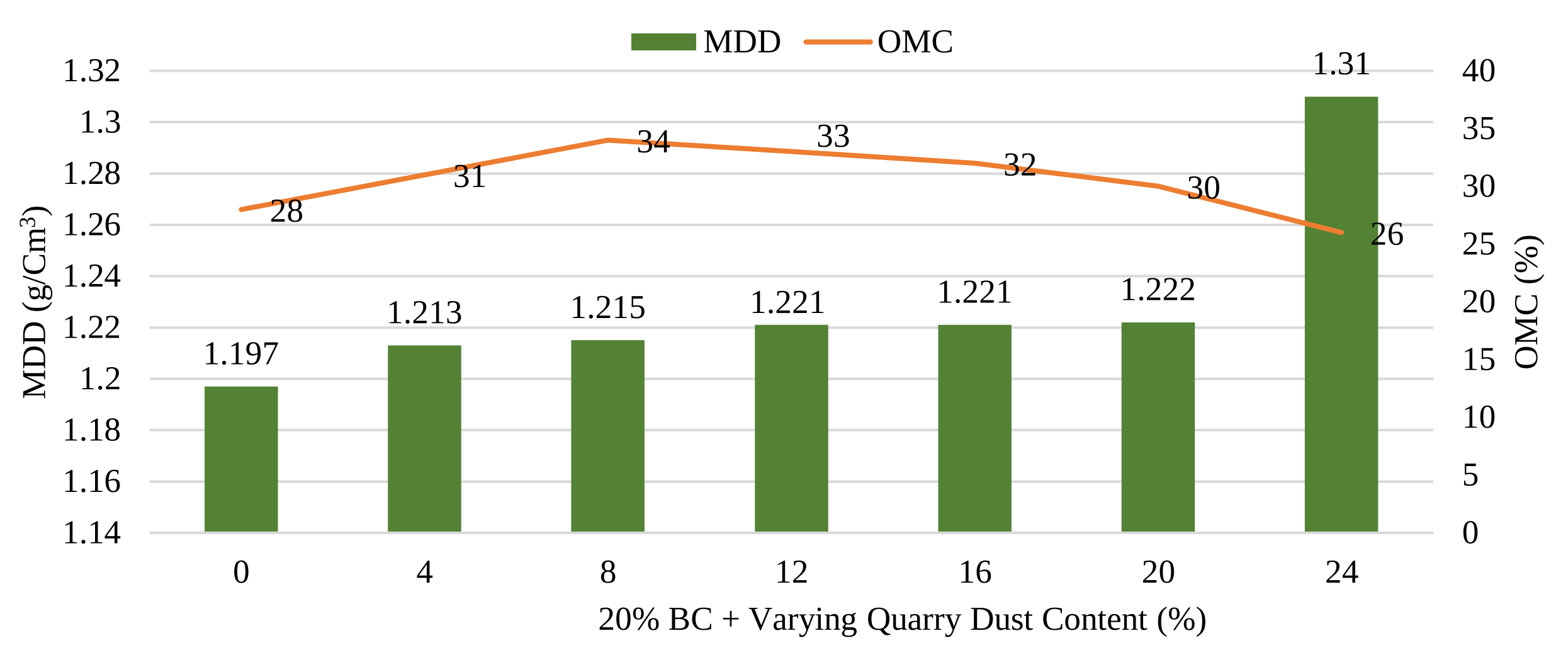

**Fig. 11 Effect of BC-QD addition on MDD and OMC of the Black Cotton Soil**

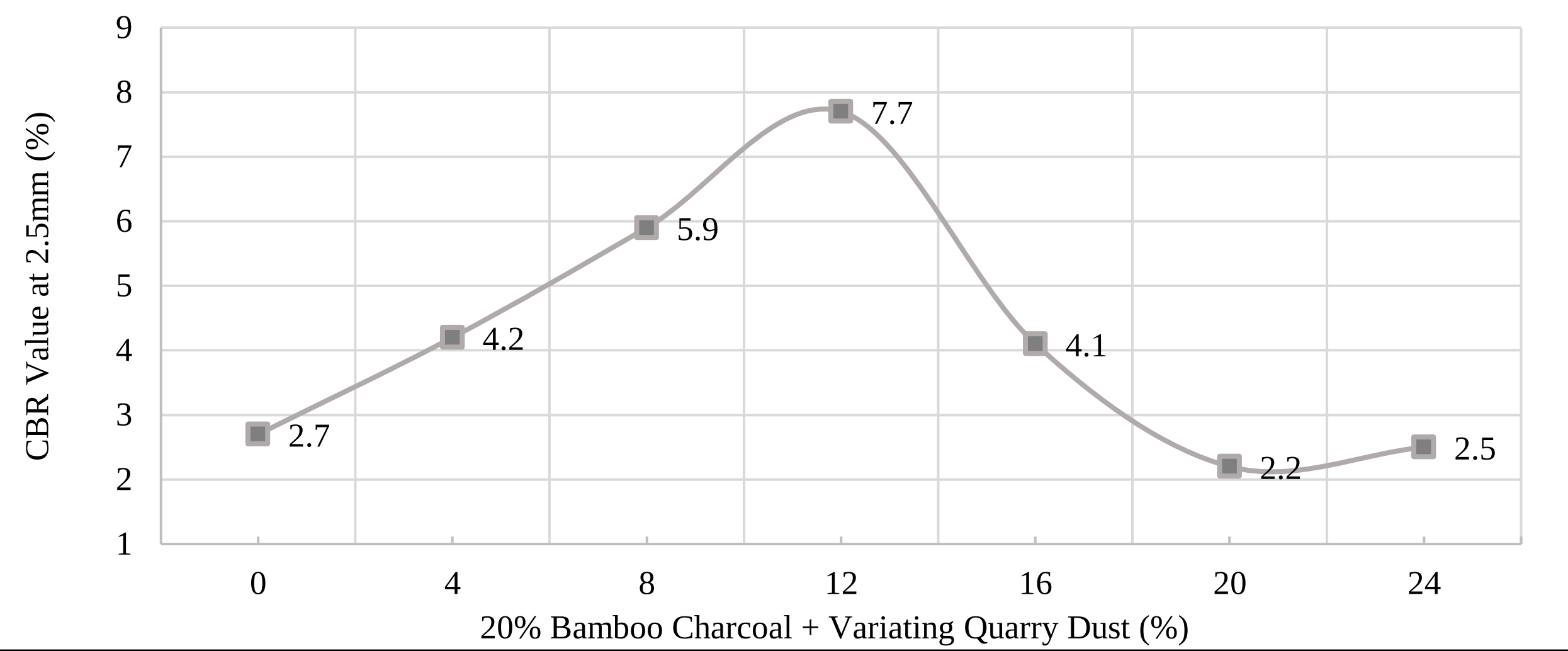

**Fig. 12 Effect of BC-QD addition on CBR value of the Black Cotton Soil**

*3.3.2. Shrinkage Limits*

Similar to the positive trend observed in Phase 1 with BC alone, adding QD further enhanced the Black Cotton Soil's volume stability. The linear shrinkage demonstrated a consistent reduction with increasing QD content. Notably, the shrinkage limit decreased from 13.25% (BCS + BC) to 12.86% at 12% QD (BCS + BC + 12% QD) (Figure 9). This improvement can be attributed to BC's porous structure [17, 18] and QD's fine particles, which likely filled voids and facilitated better water distribution within the soil matrix. This finding suggests that the stabilized soil with optimal QD content experiences less volume change during drying cycles, enhancing its suitability for subgrade applications.

*3.3.3. Free-Swell Index (FSI)*

The subsequent addition of Quarry Dust (QD) to the Black Cotton Soil (BCS) samples treated with 20% BC led to a progressive reduction in the free swell (Figure 10). Specifically, at 24% QD, the free swell decreased to 29%. This indicates that combining BC and QD further improved the soil's resistance to moisture-induced swelling. The porous structure of both BC [17] and QD [20] likely contributed to enhanced soil porosity and moisture retention properties, resulting in a more stable soil with reduced expansiveness.

*3.3.4. Compaction Characteristics (MDD and OMC)*

The combined addition of BC and QD led to intriguing changes in the soil's compaction behaviour (Figure 11). The initial addition of QD up to 8% resulted in a desirable increase in Maximum Dry Density (MDD) from 1.197 g/cm³ (BCS + BC) to 1.221 g/cm³ (BCS + BC + 8% QD). This observation suggests that the fine particles in QD effectively filled voids between soil particles, contributing to a denser and more stable soil structure. However, surpassing 8% QD content led to a noticeable decline in MDD, indicating a transition towards a lighter and less compacted state. This trend highlights the complex interplay between BC and QD, emphasizing the importance of careful optimization to achieve the optimal packing balance for subgrade applications.

*3.3.5. Strength Improvement (CBR for Four Days Soaked)*

The California Bearing Ratio (CBR) values significantly enhanced the soil's load-bearing capacity with the combined treatment. Starting from 2.7% with BC alone, the soaked CBR values progressively increased, reaching a peak of 7.7% at 12% QD content (Figure 12). This substantial improvement showcases the effectiveness of BC and QD in addressing the low-strength limitations of Black Cotton Soil. However, exceeding 12% QD resulted in a decline in CBR values, demonstrating diminishing returns and the criticality of maintaining the optimal balance between the two additives. This finding underscores the importance of precise dosage control for achieving the desired strength parameters for subgrade construction.

***3.4. Stabilization of the Black Cotton Soil, Bamboo Charcoal, and Quarry Dust with the Addition of Lime***

Building upon the promising results achieved with Bamboo Charcoal (BC) and Quarry Dust (QD) in Phase 2, Phase 3 incorporated Lime (L) to refine the soil's characteristics for subgrade applications further. The primary aim was to reduce the Plasticity Index (PI) according to [8] and enhance strength (CBR exceeding 10%) according to Kenyan standards [28]. This revision highlights critical observations and insights, emphasizing the impact of Lime on Atterberg limits, compaction behavior, and strength characteristics.

*3.4.1. Atterberg Limits and Workability*

The introduction of 5% lime significantly improved workability and reduced moisture sensitivity. Liquid Limit (LL) decreased from 66% to 59.9%, while Plastic Limit (PL) increased from 37.87% to 48.7%. This resulted in a substantial PI reduction from 28.57% (Phase 2) to 11.2%, falling well within the recommended range for subgrade materials.

Many researchers have found that adding lime to such types of soils decreases their liquid limit, increases their plastic limit, and reduces their plasticity index [8]. According to [25] findings, expansive soils lose some of their flexibility and swelling ability when 5% lime is added. This improvement signifies enhanced constructability and reduced susceptibility to deformation under moisture fluctuations.

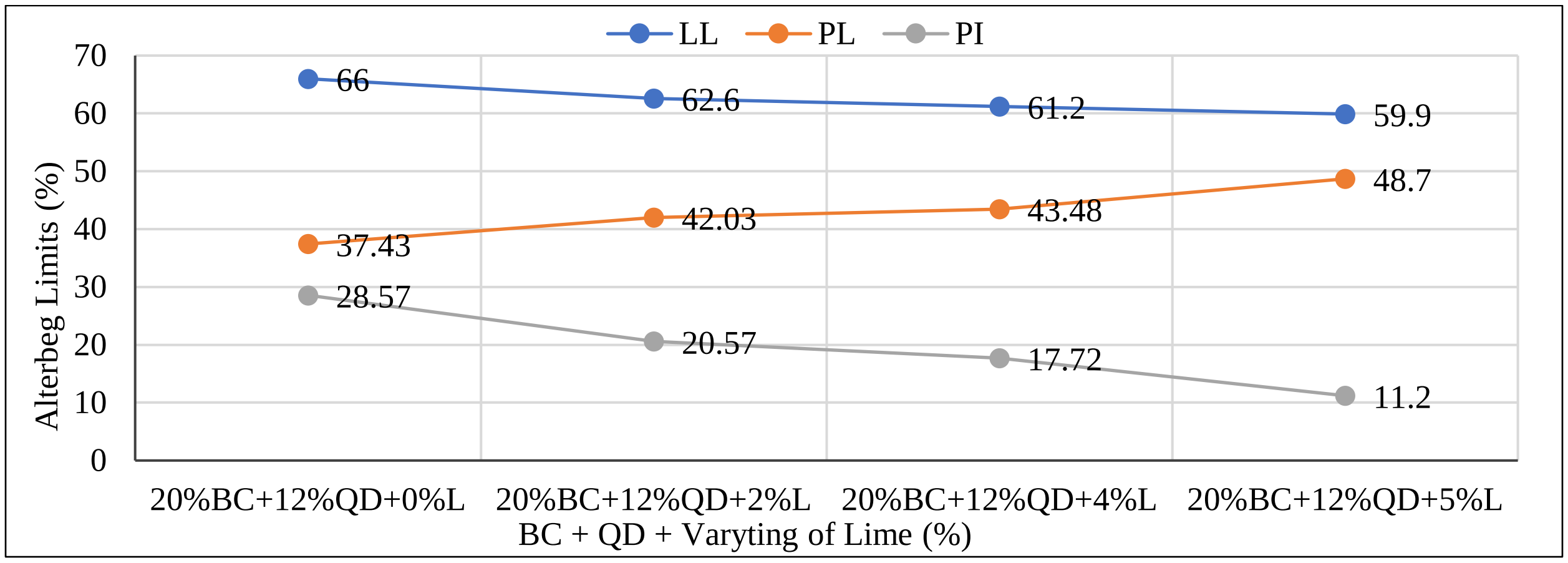

**Fig. 13 Effect of BC-QD-L addition on Atterberg limits of the Black Cotton Soil**

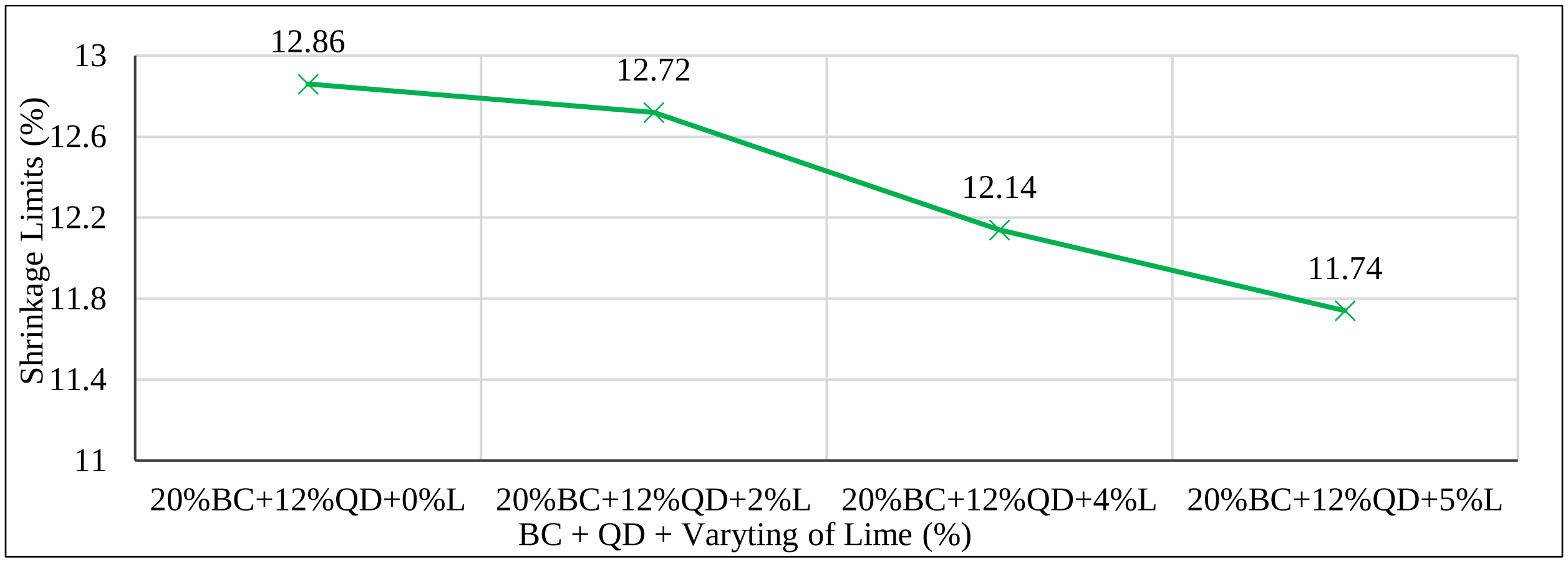

**Fig. 14 Effect of BC-QD-L addition on shrinkage limits of the Black Cotton Soil**

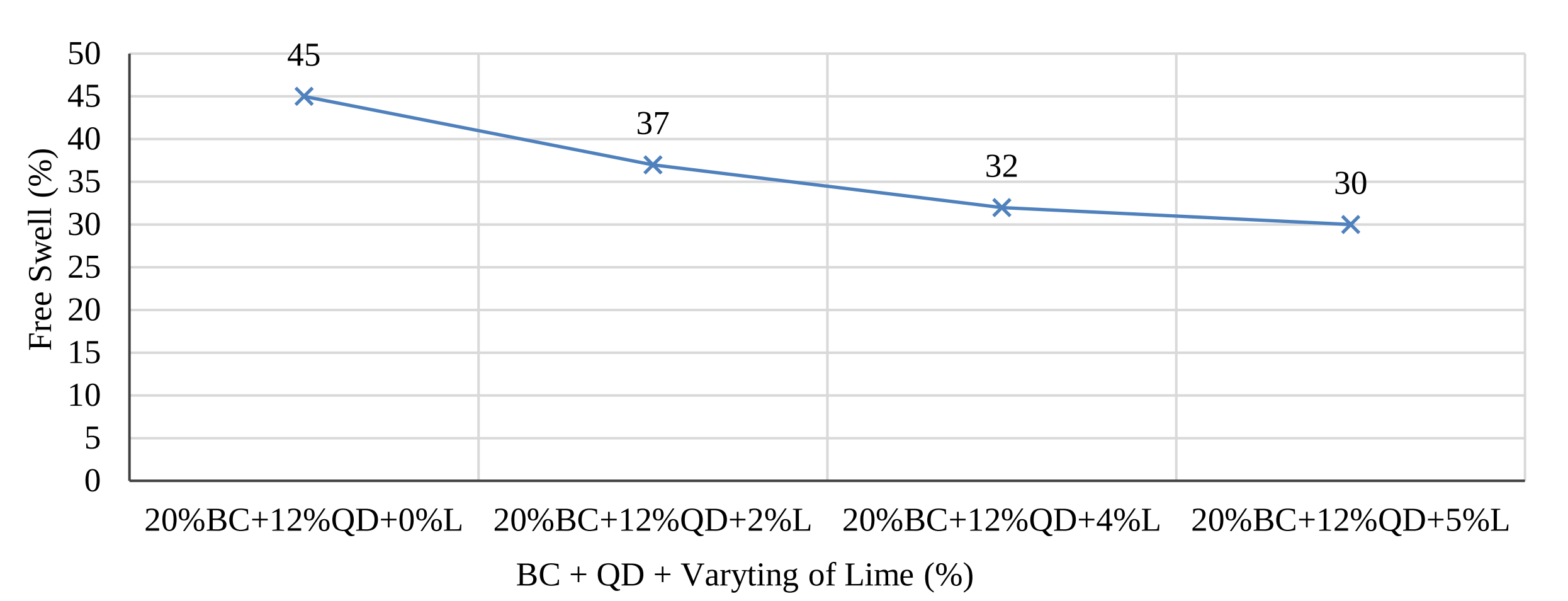

**Fig. 15 Effect of BC-QD-L addition on the Free Swell of the Black Cotton Soil**

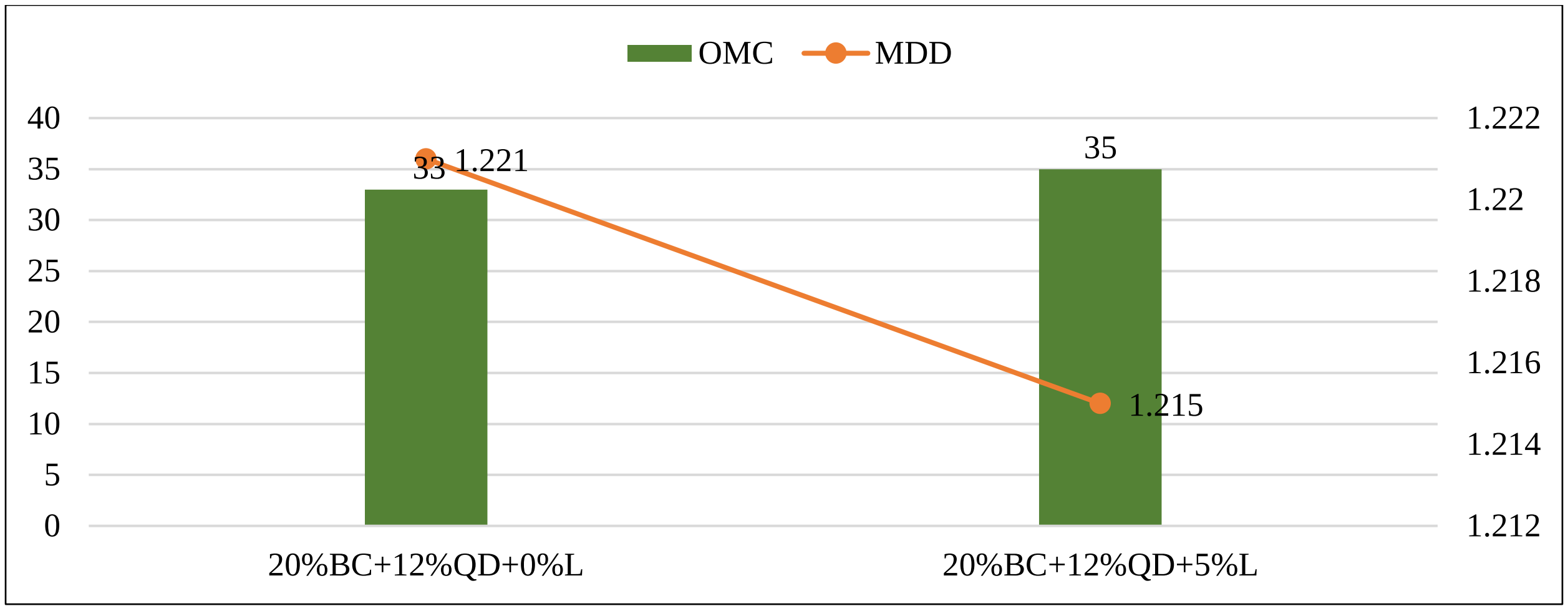

**Fig. 16 Effect of BC-QD-L addition on MDD and OMC of the Black Cotton Soil**

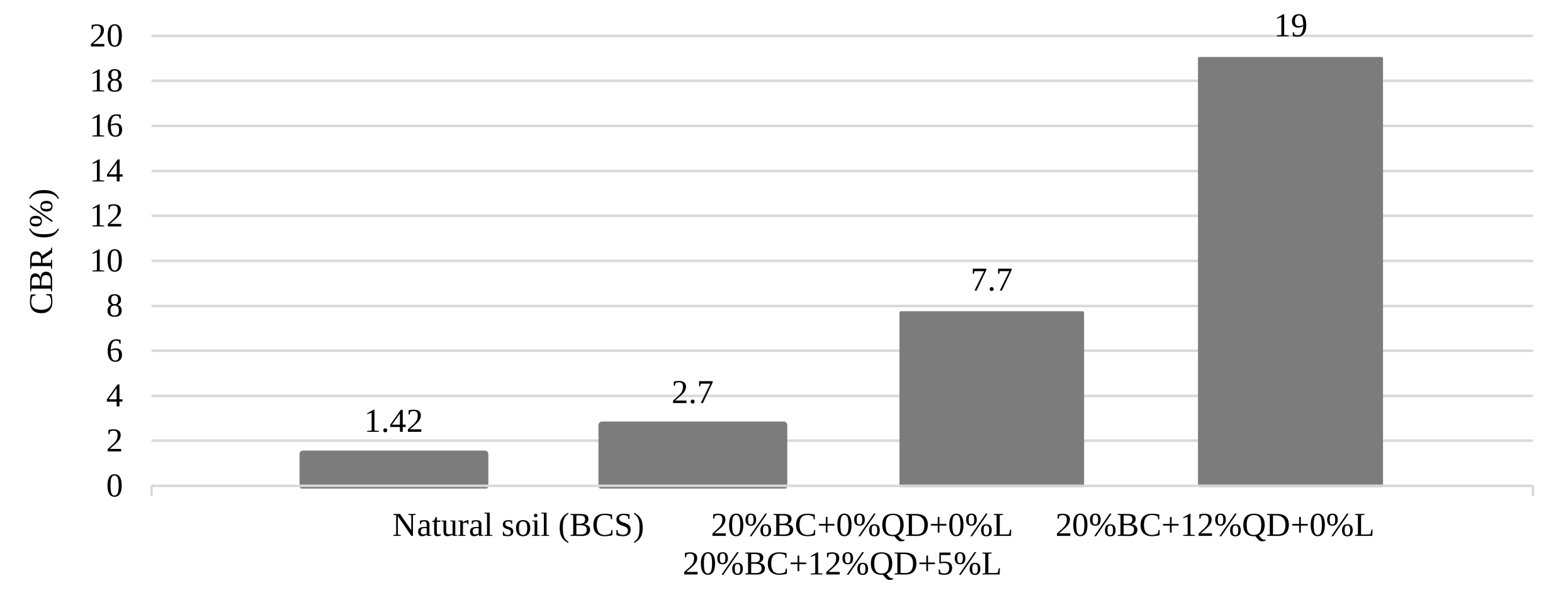

**Fig. 17 Variation of CBR values with the addition of optimum of BC, QD, and Lime**

**Table 4. Comparative analysis of the properties of untreated and treated Black Cotton Soil**

| Properties | BCS (Untreated) | BCS Treated with 20%BC | BCS Treated with 20%+12%QD | BCS Treated with 20%+12%QD+5%L |
|---|---|---|---|---|
| **Liquid Limit (LL)** | 52.26% | 66% | 66% | 59.9% |
| **Plastic Limit (PL)** | 29.42% | 39.9% | 37.43% | 48.7% |
| **Plasticity Index (PI)** | 23.14% | 26.1% | 28.57% | 11.2% |
| **Linear Shrinkage (Ls)** | 14.86% | 13.25% | 12.86% | 11.74% |
| **Free Swelling Index (FSI)** | 105% | 57% | 45% | 30% |
| **Maximum Dry Density (MDD)** | 1.388g/cm³ | 1.197g/cm³ | 1.221g/cm³ | 1.215g/cm³ |
| **Optimum Moisture Content (OMC)** | 25.5% | 28% | 33% | 35% |
| **CBR (4 Days Soaked)** | 1.42% | 2.7% | 7.7% | 19% |

*3.4.2. Shrinkage and Stability*

Shrinkage values further decreased with Lime addition, reaching a minimum of 11.74% compared to 12.86% in Phase 2. This highlights lime's contribution to enhanced soil stabilization and reduced volume changes during drying cycles, which is critical for subgrade performance. The synergy between BC, QD, and Lime effectively minimized shrinking and swelling potential, promoting long-term stability and resilience of the stabilized soil.

*3.4.3. Free-Swell Index (FSI)*

Finally, the addition of Lime (L) to the Black Cotton Soil (BCS) samples treated with 20% BC and 12% QD resulted in a further reduction in free swell (Figure 15). Specifically, at 5% L, the free swell index decreased to 30%. This indicates that the combination of BC, QD, and L contributed to an even more significant improvement in the soil's resistance to moisture-induced swelling. The porous structure of BC [17] and QD, combined with the stabilizing effect of L, likely enhanced soil porosity, moisture retention properties, and overall soil stability.

*3.4.4. Compaction Characteristics*

While the Maximum Dry Density (MDD) decreased slightly from 1.221 g/cm³ (Phase 2) to 1.215 g/cm³ with 5% Lime, the change was marginal. This suggests that the soil structure remained dense enough for subgrade construction. Conversely, lime increased the Optimum Moisture Content (OMC) from 33% to 35%. This indicates a slightly higher moisture requirement for optimal compaction, likely due to lime's water-binding properties influencing the soil-moisture interaction.

*3.4.5. Strength Enhancement and Subgrade Suitability*

The California Bearing Ratio (CBR) value remarkably increased, peaking at 19% with the combined treatment of 20% BC, 12% QD, and 5% Lime. This represents a significant improvement compared to Phase 1 (2.7%) and Phase 2 (7.7%), achieving the desired strength required for subgrade applications. The combined effect of BC, QD, and Lime demonstrates a synergistic relationship in enhancing the soil's load-bearing capacity, meeting the prescribed CBR standards outlined in the Kenya Roads Design Manual [28].

Adding 5% lime proved to be the optimal dosage for this specific mixture, successfully achieving the desired plasticity, shrinkage, and strength parameters while remaining within the recommended Lime range for construction purposes. This phase conclusively demonstrates the effectiveness of utilizing BC, QD, and Lime in conjunction to stabilise Black Cotton Soil and transform it into a material suitable for road subgrade construction.

## 4. Microstructural Properties Analysis

This section aims to utilize Scanning Electron Microscopy (SEM) to examine and elucidate the structural modifications induced by each stabilizing agent in the composite material. Through a series of SEM images, the section aims to provide a detailed understanding of the evolving microstructure at each modification stage, correlating these changes with observed alterations in soil properties such as plasticity, compaction, and mechanical strength. The ultimate goal is to establish a comprehensive link between microstructural transformations and the enhanced performance of the stabilized soil, particularly in the context of road subgrade material.

Scanning Electron Microscopy (SEM) in Figure 18 provided a profound insight into the microstructural changes induced by each stabilizing agent. Initially, the SEM imagery of untreated BCS (a) revealed a loose and heterogeneous matrix prone to plastic deformation and swelling. Adding 20% Bamboo Charcoal (b) altered this matrix into a more porous structure, likely contributing to enhanced particle interlock and reduced soil plasticity. Subsequent incorporation of 12% Quarry Dust (c) resulted in a denser packing of soil particles, as evidenced by the SEM, potentially leading to improved compaction (Figure 11) and mechanical strength (Figure 12).

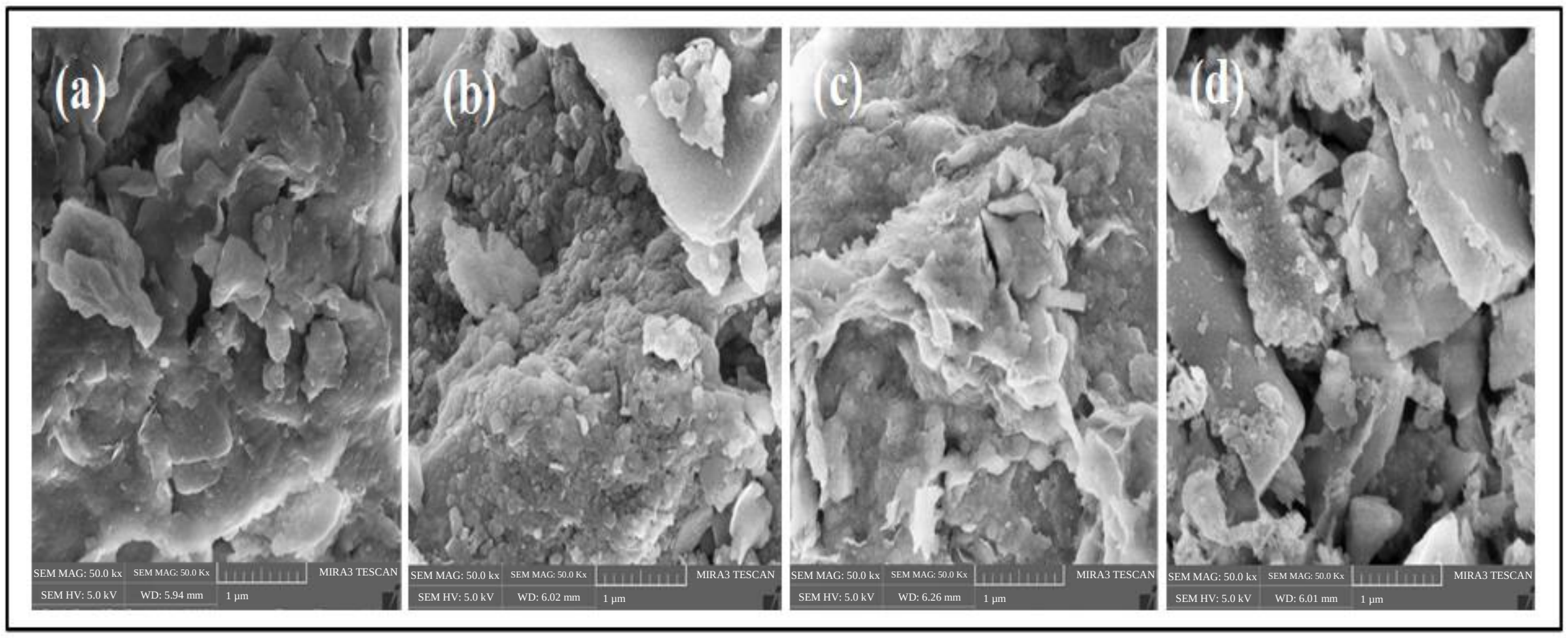

**Fig. 18 (a) SEM microphotograph of BCS, (b) BCS+20%BC, (c) BCS+20%BC+12%QD, and (d) BCS+20%BC+12%QD+5%L .**

Finally, the addition of 5% Lime (d) created a highly cohesive and compacted microstructure, with SEM images suggesting the formation of cementitious bonds, correlating with the highest CBR values observed in Figure 17 and indicating a substantial increase in soil stability suitable for road subgrade material (Table 4).

## 5. Conclusion

This study meticulously tackled Black Cotton Soil's high plasticity, considerably high swell index, and poor load-bearing capacity, major hindrances in road subgrade construction. Through a triphasic stabilization strategy involving Bamboo Charcoal, Quarry Dust, and Lime, we achieved: A substantial increase in CBR from 1.42% to 19%, exceeding the Kenya Roads Design Manual Part III requirements. The Plasticity Index was significantly reduced from 23.14% to a manageable 11.2%, enhancing workability and resilience against moisture-induced damage. Overall compaction characteristics were substantially enhanced, providing a foundation for an efficient and durable subgrade.

The mitigation of issues related to cracked and deformed roads became evident, with swelling index and shrinkage brought within desirable limits. The SEM analysis confirmed microstructural transformations that underpin these improvements, presenting a sustainable solution for enhancing the geotechnical properties of expansive soils. These findings provide a practical avenue for rural road development and contribute to the sustainable management of problematic soils.

## Funding Statement

This research was funded by the African Union Commission (AUC).

## Acknowledgments

The Jomo Kenyatta University of Agriculture and Technology (JKUAT) in Nairobi, Kenya, and the Pan African University Institute for Basic Sciences, Technology, and Innovation (PAUSTI) for their assistance and provision of essential resources are gratefully acknowledged by the researchers.